\documentclass[
aps,prd,
12pt,
nopreprintnumbers,
showpacs,
eqsecnum,
nofootinbib
]{revtex4-1}

\usepackage{graphicx}

\def\Tr{{\rm Tr\,}}

\def\v0{{\bf 0}}

\begin{document}

\title{
Nambu-Jona-Lasinio Model and Deconstructed Dimension}
\author{Nahomi Kan}\email[]{kan@yamaguchi-jc.ac.jp}
\affiliation{
Yamaguchi Junior College,
Hofu-shi, Yamaguchi 747--1232, Japan}
\author{Koichiro Kobayashi}\email[]{m004wa@yamaguchi-u.ac.jp}
\author{Kiyoshi Shiraishi}\email[]{shiraish@yamaguchi-u.ac.jp}
\affiliation{
Yamaguchi University,
Yamaguchi-shi, Yamaguchi 753--8512, Japan}
\date{\today}

\begin{abstract}
The Nambu-Jona-Lasinio 
model with the mass matrix which appears in a 
deconstruction model is investigated. We consider two models.
In Model A, a mass matrix belonging to  a type used in dimensional
deconstruction is introduced. In Model B, the four-fermion interaction
has a structure of the matrix of the type of dimensional deconstruction.
In these models, we find that generation of a dynamical fermion mass
spectrum occurs in a strong coupling case.
\end{abstract}


\pacs{
11.10.Kk, 
11.25.Mj, 
11.30.Rd, 
11.30.Qc, 
}

\maketitle

\section{Introduction}
The Nambu-Jona-Lasinio (NJL) model \cite{NJL}, which
consists only of fermions, is still important as a model of
dynamical symmetry breaking considered in theories beyond the standard
model.  Recently, Bunk et al. have studied the NJL
model in spacetime with an extra dimension \cite{bhst}. 
As in a non-renormalizable model, it is worth studying the effect of
two different scales, the cutoff scale and the compactification scale.

In the present paper,
we introduce the four-dimensional models similarly as in the view point of
dimensional deconstruction \cite{Deconstruction} and investigate the
one-loop quantum effect in a theory with a finite spectrum of fermions.
In dimensional deconstruction models \cite{Deconstruction}, excitation
modes exist, similarly in higher-dimensional theory, and it is known that 
similar features are found as in higher-dimensional theory in a certain
case. In a model which consists of four dimensional fermions
of a limited number as in the dimensional deconstruction, we wish
to investigate how quantum effects bring about symmetry breaking.

Two situations
can be considered about chiral symmetry in construction of our
models.  One is the case where it has a
mass spectrum based on a deconstruction model. In this case,
quantum correction to the fermion mass would be observed and 
a large amount of condensation of fermion is expected for a certain strong
coupling. Another situation is the case where chiral symmetry is imposed
about all the fermion fields in the model. In this case, the
condensation of fermions through quantum effect at strong coupling
produces massive states of fermions. Even in this case, as seen later, the
technique used in a deconstruction model is valid for obtaining the
one-loop effect by a prescription using auxiliary fields.

The plan of this paper is as follows.
In the next section, we consider Model A, in which the mass matrix of
fermions is introduced. Similarity to a five dimensional
four-fermion model will be discussed.
In Sec. \ref{sec3}, we consider  a massless NJL model with
`non-diagonal' four-fermion interactions, as Model B. The interacting term
is chosen as the effective fermion masses have a spectrum akin to that in
a deconstruction model.
We end with a summary in Sec. \ref{sec4}.

\section{model A}
In this section, we consider a type of a massive NJL model.
The free part of the Lagrangian for fermion fields takes the form
\begin{equation}
{\cal L}_f=\sum_{k=1}^N\left[i\bar{\lambda}_{k}\partial\!\!\!/\lambda_{k}+
i\bar{\chi}_{k}\partial\!\!\!/\chi_{k}
\right]
+\sum_{k=1}^N \left[\bar{\lambda}_{k}(m_0\chi_{k}-f\chi_{k+1})+{\rm
h.c.}\right]
\,,
\label{free}
\end{equation}
where $\lambda_k$ and $\chi_k$ are Weyl spinors with opposite
chiralities. The `bare' mass
$m_0$ and the constant
$f$ have dimension of mass and take real values. The $N+1$-st fields
should be identified to the first one, i.e., $\chi_{N+1}\equiv\chi_1$. The
eigenvalues for the fermion mass-squared are given by
\begin{equation}
\bar{M}_p^2=m_0^2+f^2-2f m_0 \cos\frac{2\pi p}{N}\,,
\end{equation}
with $p=0, 1, \dots, N-1$ and if and only if $m_0=f$, a zero eigenvalue
exists
\cite{KSJMP}.
The eigenmodes turn out to be
\begin{equation}
\chi^{(p)}=\frac{1}{\sqrt{N}}\left(
\begin{array}{c}
1  \\
e^{i\frac{2\pi p}{N}} \\
e^{i\frac{4\pi p}{N}} \\
\vdots \\
e^{i\frac{2\pi (N-1)p}{N}} 
\end{array}
\right)\,.
\end{equation}

Now, we introduce an interacting part of the Lagrangian.
We assume it as
\begin{equation}
{\cal L}_G=G\sum_{k=1}^N
(\bar{\lambda}_k\chi_k)(\bar{\chi}_k\lambda_k)\,,
\end{equation}
where $G$ is the four-fermion coupling
and the total Lagrangian is supposed to be ${\cal L}={\cal L}_f+{\cal
L}_G$.
The four-fermi interaction is `local' with respect to each entry of
the fermion, and it is different from the interaction in the Gross-Neveu
model \cite{GN}. Apparently, if $m_0=f$, only the zero-mode fermion
possesses chiral symmetry.

The interaction term can be replaced by the term including an auxiliary
field. That is, the interacting part of the Lagrangian can be rewritten to
\begin{equation}
{\cal L}'_G=
-m_0\sum_{k=1}^N (\bar{\lambda}_{k}\chi_{k}+\bar{\chi}_{k}\lambda_{k})+
\sum_{k=1}^N\left[
(\bar{\lambda}_k\chi_k)\sigma_k+(\bar{\chi}_k\lambda_k)\sigma_k^*\right]
-\sum_{k=1}^N\frac{|\sigma_k-m_0|^2}{G}\,,
\end{equation}
where the complex fields $\sigma_k$ ($k=1, 2, \dots, N$) are auxiliary
fields.

Throughout the present paper, we use a homogeneity ansatz
\begin{equation}
\sigma_1=\sigma_2=\cdots=\sigma_N\equiv\sigma\,,
\end{equation}
and the modulation with respect to $k$ is assumed to be negligible.
Now, we can obtain the effective potential for
$\sigma$ up to one-loop order  in the following form:
\begin{equation}
V(\sigma)=\frac{N|\sigma-m_0|^2}{G}+i\sum_{p=0}^{N-1}\Tr\ln
(i\partial\!\!\!/+M_p)\,,
\end{equation}
where the effective mass of fermions are given by
\begin{equation}
M_p^2=\left|\sigma-f e^{i\frac{2\pi p}{N}}\right|^2=|\sigma|^2+f^2-2f
|\sigma| \cos\vartheta_p\,,
\end{equation}
where
$\vartheta_p=\frac{2\pi p}{N}-\arg (\sigma)$.
The amount of fermion condensation is determined by the gap equation
$\frac{dV}{d\sigma}(\langle\sigma\rangle)=0$.

The one-loop effective potential is evaluated formally by Schwinger's
parameter integral as
\begin{equation}
2\int\frac{d^4k}{(2\pi)^4}\int_{0}^\infty\frac{dt}{t}
\sum_{p=0}^{N-1}\exp\left[-(
k^2+M_p^2)t\right]
=\frac{2}{(4\pi)^2}\int_0^\infty\frac{dt}{t^3}\sum_{p=0}^{N-1}
\exp\left[-M_p^2t\right]
\,.
\end{equation}
Furthermore, we can apply the formula \cite{KSS}
\begin{equation}
e^{(2f^2\cos\vartheta) t}=\sum_{\ell=-\infty}^\infty
e^{i\ell\vartheta}I_\ell(2f^2t)\,,
\end{equation}
where $I_\ell(z)$ is the modified Bessel function,
to the expression. Then we find

\begin{equation}
V(\sigma)=\frac{N|\sigma-m_0|^2}{G}+\frac{2}{(4\pi)^2}\sum_{p=0}^{N-1}\sum_{\ell=-\infty}^\infty
e^{i\ell\vartheta_p}{\cal I}(\ell;f,|\sigma|)\,,
\end{equation}
where
\begin{equation}
{\cal
I}(\ell;f,|\sigma|)\equiv\int_0^\infty\frac{dt}{t^3}\exp
\left[-(f^2+|\sigma|^2)t\right]I_\ell(2f|\sigma|t)\,.
\end{equation}
Now, the summation over $p$ can be performed, and it turns out to be
\begin{equation}
V(\sigma)=\frac{N|\sigma-m_0|^2}{G}+NV_1(\sigma)\,,
\end{equation}
with
\begin{eqnarray}
&
&V_1(\sigma)=\frac{2}{(4\pi)^2}\sum_{q=-\infty}^\infty
e^{-iqN\arg (\sigma)}{\cal I}(qN;f,|\sigma|)\nonumber \\
& &=\frac{2}{(4\pi)^2} {\cal
I}(0;f,|\sigma|)+\frac{4}{(4\pi)^2}\sum_{q=1}^\infty
\cos[qN\arg (\sigma)]{\cal I}(qN;f,|\sigma|)\,.
\end{eqnarray}

The integral ${\cal I}(0;f,|\sigma|)$ is  divergent  as in the
present form.%
\footnote{Note that ${\cal I}(\ell;f,|\sigma|)$ for $\ell\ge 3$ is
finite.}  
Therefore, some regularization is needed.
Several regularization methods on Schwinger's parameter integral 
have been considered by many authors.
A `minimal subtraction' method \cite{EGOS} uses the cutoff method only
for the divergent part as in the original NJL model.
A use of density with cutoff \cite{HOBP} is also considered.
In the present paper, we adopt the simple introduction of the cutoff
scale $\Lambda$, as for the lower bound of $t$ as
$\Lambda^{-2}$ \cite{IKT}.
Namely, we use ${\cal I}_\Lambda (\ell;f,|\sigma|)$ defined by
\begin{equation}
{\cal
I}_\Lambda(\ell;f,|\sigma|)\equiv\int_{\Lambda^{-2}}^\infty\frac{dt}{t^3}\exp
\left[-(f^2+|\sigma|^2)t\right]I_\ell(2f|\sigma|t)\,,
\end{equation}
instead of ${\cal
I}(\ell;f,|\sigma|)$.

\begin{figure}[h]
\centering
\includegraphics
{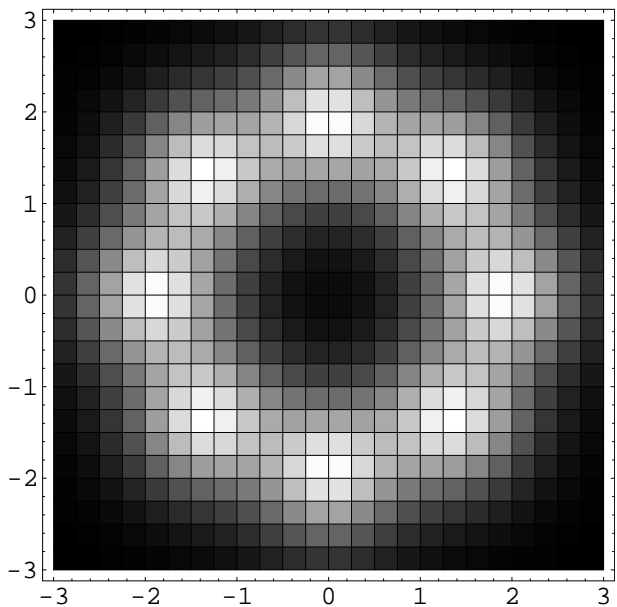}
\caption{%
A density plot of $V_1(\sigma)$ for $N=8$ and $f/\Lambda=2$.
The axes indicate the real and imaginary part of
$\sigma$.}
\label{fig1}
\end{figure}

It is obvious that the one-loop part of the potential, $V_1(\sigma)$, has
a
$Z_N$ symmetry;
$V_1(\sigma)$ is invariant under the change that $\arg(\sigma)\rightarrow
\arg(\sigma)+\frac{2\pi}{N}$.
A typical shape is shown in Fig. \ref{fig1} as in a density plot, where
the brighter region means the larger value of the potential.

We find that the maximum of $V_1$ is located at the value for $|\sigma|$
which is smaller than
$f$. Thus, if the bare mass $m_0=f$,  the `dynamical' mass
$m=\langle |\sigma|\rangle$ becomes larger than $f$, unless $G=0$.
This is due to the interaction term which mixes the eigenmodes of the free
theory.

\begin{figure}[h]
\centering
\includegraphics
{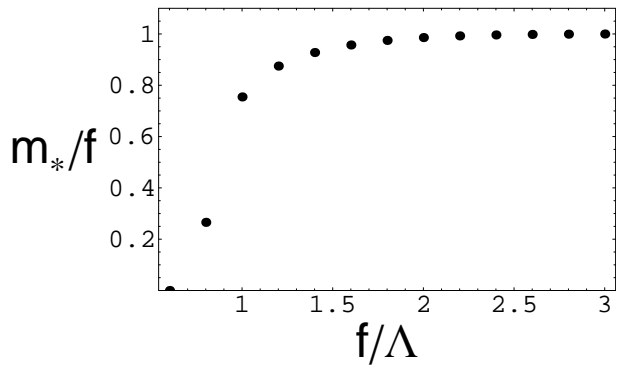}
\caption{%
$m_*/f$ is plotted against $f/\Lambda$ for $N=8$.
}
\label{fig2}
\end{figure}

Conversely, the choice of $m_0$ can lead to the unmodified mass, $m=m_0$,
for any small value for $G$. This can be achieved by setting
the maximum of $V_1$ is located at $\sigma=m_0$. Then the condensation of
fermions vanishes. In this case, let us denote the special value for mass
as
$m_*$, which is expressed as a function of $f/\Lambda$. In Fig.
\ref{fig2},
$m_*/f$  is plotted against
$f/\Lambda$. 
For large $f/\Lambda$, the value of $m_*$
approaches to the value $f$.

Incidentally, we recognize an existence of the case that 
$m=f$ is possible for the bare mass in the range
$m_*<m_0<f$ with an appropriate tuning of $G$. Thus, when $f/\Lambda$
becomes large,
$(m-m_0)/f$ converges to zero also in this case.
The qualitative features stated so far are independent of $N$.

\begin{figure}[h]
\centering
\includegraphics
{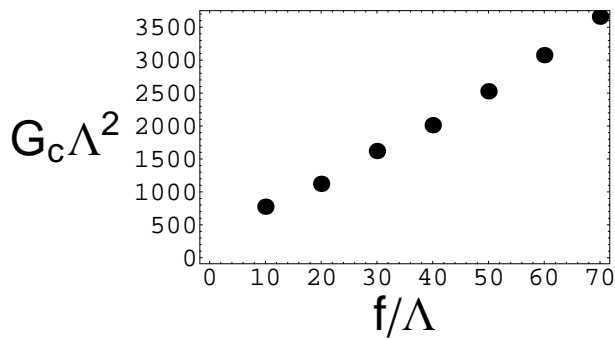}
\caption{%
$G_c\Lambda^2$ in the case of $m_0=m_*$ is plotted against $f/\Lambda$ for
$N=8$. }
\label{fig3}
\end{figure}

The critical coupling $G=G_c$ for fermion condensation is calculated from
the second derivative $V(\sigma)$ at $m=m_*$. If $G>G_c$, the effective
potential $V(\sigma)$ becomes a local maximum at $\sigma=m_*$. Then, 
condensation of fermions is expected to pick a large value and even the
lowest mass eigenvalue of fermions acquires a large mass.  We show the
value of
$G_c\Lambda^2$ for $m_0=m_*$ in Fig.
\ref{fig3}, for relatively large values of
$f/\Lambda$.

Now, we further investigate the limiting case that
$f/\Lambda$ is large.
For $N\gg 1$ and $f|\sigma|\Lambda^{-2}\gg 1$,
we can estimate the one-loop effect as
\begin{eqnarray}
V_1(\sigma)&\approx&\frac{2\Lambda^4}{(4\pi)^2}\int_1^\infty\frac{dt}{t^3}\exp
\left[-(f^2+|\sigma|^2)\Lambda^{-2}t\right]I_0(2f|\sigma|\Lambda^{-2}t)\nonumber
\\ &\approx&\frac{2\Lambda^4}{(4\pi)^2}\int_1^\infty\frac{dt}{t^3}\exp
\left[-(f-|\sigma|)^2\Lambda^{-2}t\right]
\frac{1}{\sqrt{4\pi f|\sigma|\Lambda^{-2}t}}\,.
\end{eqnarray}
Therefore in the region $\sigma\approx f$, $V_1$ is approximated as
\begin{equation}
V_1(\sigma)\approx\frac{2\Lambda^5}{(4\pi)^{5/2}f}\int_1^\infty
\frac{dt}{t^{7/2}}\exp
\left[-(f-|\sigma|)^2\Lambda^{-2}t\right]\,.
\end{equation}
The critical value for $G_c$ calculated from the approximation
appears to be proportional to $f$. This observation qualitatively agrees
with the result shown in Fig.
\ref{fig3}.

The comparison to a higher-dimensional model is worth studying.
Let us consider a five-dimensional interacting four-fermion model
governed by the Lagrangian
\begin{equation}
{\cal
L}_{KK}=i\bar{\psi}{\partial\!\!\!/}_{(5)}\psi+
G_5\bar{\psi}\psi\bar{\psi}\psi\,,
\end{equation}
where ${\partial\!\!\!/}_{(5)}$ is the Dirac operator defined in five
dimensions. By using a real auxiliary field $S$, we can rewrite this
as%
\footnote{As is well known, the quality of approximation is
guaranteed for large $N$ fermion models. We consider here a simple model
for explanation.}
\begin{equation}
{\cal
L}_{KK}=i\bar{\psi}{\partial\!\!\!/}_{(5)}\psi+S\bar{\psi}\psi-
\frac{S^2}{4G_5}\,.
\end{equation}
Suppose that the fifth dimension is compactified to a circle whose
circumference is $L$. Then, the one-loop effective potential for the
auxiliary field $S$ is given by
\begin{equation}
V_{KK}(\sigma)=\frac{S^2}{4G_5}+i\frac{1}{L}\sum_{p=-\infty}^{\infty}
\Tr\ln
(i\partial\!\!\!/+M_p)\,,
\end{equation}
where $M_p=S+i\gamma_5\frac{2\pi p}{L}$.

For large value of $L\Lambda$, this expression can be approximated,
using an identity for the elliptic theta function, as
\begin{equation}
V_{KK}(\sigma)\approx\frac{S^2}{4G_5}+\frac{2\Lambda^5}{(4\pi)^{5/2}}
\int_1^\infty \frac{dt}{t^{7/2}}\exp\left[-\frac{S^2}{\Lambda^2}
t\right]\,.
\end{equation}
By comparison of two models, the critical coupling $G_{5c}$ for 
condensation of the fermion in the five dimensional model is found
to be related to the critical coupling
$G_c$ in the four dimensional model as the following manner:
\begin{equation}
{G_{5c}}\approx\frac{G_c}{4f}\,,
\end{equation}
for large $f/\Lambda$.
Because the relation between deconstruction models and Kaluza-Klein models
is $L=N/f$ for large $f$ and large $N$ \cite{Deconstruction,KSJMP},
the relation reads
$G_{c5}/L\approx G_c/N$. The appearance of the characteristic number of
the model,
$N$, indicates that the five-dimensional model and the four-dimensional
model considered here have a common feature but are not entirely
equivalent in the limit.
{The dimensional deconstruction has a continuum limit to the
higher-dimensional theory for free fields, but exact discretization is
not realized by the method generally.}

The difference in the degree of freedom of auxiliary fields between four
and five dimensional model is due to the absence of
chirality in five dimensions.
We think that more similar five-dimensional model is the model with
a $U(1)$ gauge field.
 The vector field in the compactified fifth dimension
plays a role of the imaginary part of the order parameter.
The analysis on such a model will be presented elsewhere.

\section{model B\label{sec3}}

The model B proposed in this section does not include mass scales as input
parameters.
This model mimics the model A, in view of the coupling between adjacent
fermions. Simply speaking, the mass scale $f$ is supposed to be replaced
by an auxiliary field $\rho_k$.
The Lagrangian of the model B consists of the free part (\ref{free}) with
$m_0=f=0$ and the interacting part, expressed by use of auxiliary fields,
\begin{eqnarray}
{\cal L}'_G&=&\sum_{k=1}^N\left[
(\bar{\lambda}_k\chi_k)\sigma_k+{\rm h. c. }\right]
-\sum_{k=1}^N\left[
(\bar{\lambda}_k\chi_{k+1})\rho_k+{\rm h. c. }\right]\nonumber \\
&
&-\sum_{k=1}^N\frac{\cosh\theta|\sigma_k|^2-2\sinh\theta
{\rm Re}(\sigma_k\rho_k^*)+\cosh\theta|\rho_k|^2}{G}\,,
\end{eqnarray}
where $G$ and $\theta$ are constant.
If the auxiliary fields are integrated out,
the following four-fermion interaction can be found:
\begin{eqnarray}
{\cal L}_G&=&G \sum_{k=1}^N\left[
\cosh\theta(\bar{\lambda}_k\chi_k)(\bar{\chi}_k\lambda_k)
-\sinh\theta[(\bar{\lambda}_k\chi_{k+1})(\bar{\chi}_{k}\lambda_k)
+(\bar{\lambda}_k\chi_{k})(\bar{\chi}_{k+1}\lambda_k)]\right.\nonumber \\
& &\qquad+\left.
\cosh\theta(\bar{\lambda}_k\chi_{k+1})(\bar{\chi}_{k+1}\lambda_k)
\right]\,.
\end{eqnarray}

As in the previous section, we require a homogeneity 
ansatz,
\begin{equation}
\sigma_1=\sigma_2=\cdots=\sigma_N\equiv\sigma \qquad
{\rm and}\qquad\rho_1=\rho_2=\cdots=\rho_N\equiv\rho\,.
\end{equation}
Then, the fermion mass-squared spectrum is given by
\begin{equation}
M^2_p=\left|\sigma-\rho\, e^{i\frac{2\pi p}{N}}\right|^2
=|\sigma|^2+|\rho|^2-2 |\sigma| |\rho|\cos\vartheta_p\,,
\end{equation}
where
$\vartheta_p=\frac{2\pi p}{N}+\arg (\rho)-\arg (\sigma)\equiv\frac{2\pi
p}{N}+\varphi$ and $p=0, 1, \dots, N-1$.%
\footnote{Thus, we know that $\arg (\rho)+\arg (\sigma)$ is the
Nambu-Goldstone mode.}

The effective potential in this model is expressed as
\begin{equation}
V(|\sigma|,|\rho|,\varphi)=N\frac{\cosh\theta|\sigma|^2-2\sinh\theta
\cos\varphi
|\sigma||\rho|+\cosh\theta|\rho|^2}{G}+NV_1(|\sigma|,|\rho|,\varphi)\,,
\end{equation}
with
\begin{equation}
V_1(|\sigma|,|\rho|,\varphi)=\frac{2}{(4\pi)^2} {\cal
I}(0;|\rho|,|\sigma|)+\frac{4}{(4\pi)^2}\sum_{q=1}^\infty
\cos[qN\varphi]{\cal I}(qN;|\rho|,|\sigma|)\,.
\end{equation}

\begin{figure}[h]
\centering
\includegraphics
{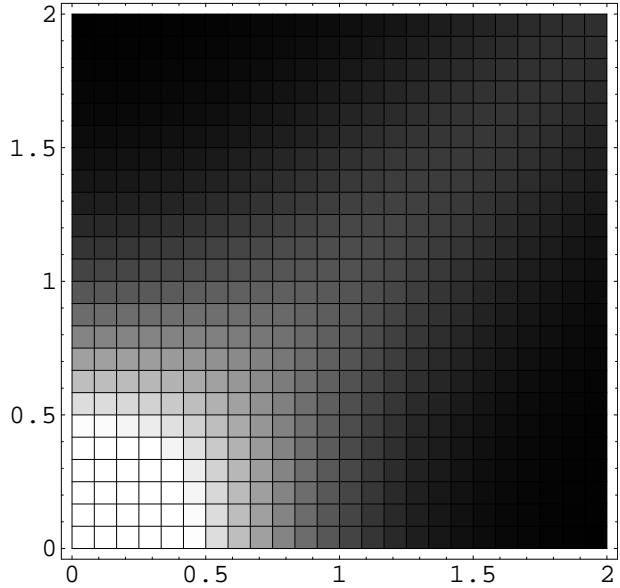}
\caption{%
$V_1(|\sigma|,|\rho|,0)$ is shown as a density plot for
$N=8$. The axes indicate $|\sigma|$ and $|\rho|$.}
\label{fig4}
\end{figure}

\begin{figure}[h]
\centering
\includegraphics
{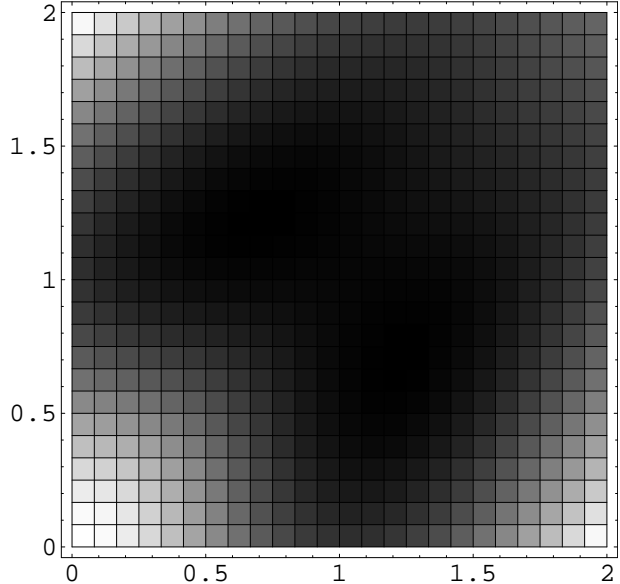}
\caption{%
$V(|\sigma|,|\rho|,0)$ is shown as a density plot for
$G\Lambda^{2}=1000$, $\theta=1$ and $N=8$. The axes indicate
$|\sigma|$ and
$|\rho|$.}
\label{fig5}
\end{figure}

In this paper, we assume $\theta>0$ and $\langle\varphi\rangle=0$
(unbroken CP).
In Fig. \ref{fig4}, we show the one-loop contribution $V_1(|\sigma|,
|\rho|, 0)$ as a density plot for $N=8$.
In Fig. \ref{fig5}, we show the effective potential  $V(|\sigma|,
|\rho|, 0)$ for the case with $G\Lambda^{2}=1000$, $\theta=1$ and $N=8$.
We find that the origin is a local maximum and the minimum point exists
evidently in this case. Thus chiral symmetry is
broken, because $\langle|\sigma|\rangle=\langle|\rho|\rangle\ne 0$ 
and  condensation of fermions occurs.
Then, the effective fermion mass spectrum is dynamically generated as
\begin{equation}
{M}_p^2=4\langle|\sigma|\rangle^2 \sin^2\frac{\pi p}{N}\,.
\end{equation}

In this model, the critical coupling for condensation can be analytically
evaluated as
\begin{equation}
G_c\, e^{\theta}\Lambda^2=8\pi^2 \,,
\end{equation}
and is found to be independent of $N$.

\section{Summary and outlook\label{sec4}}
In the present paper, we have considered two unorthodox extensions of the
Abelian NJL model.
The model A turns out to be related to a higher-dimensional model for a
large mass scale $f/\Lambda\gg 1$.
We find that the model B can exhibit a fermion mass spectrum as in the
deconstruction models.

Because of the variety of the effective fermion mass, the effective
potential at finite temperature or density may lead to various phases for
condensation. The thermodynamic properties of the models should be studied
also in cosmological models.

The models in this paper include self-interactions and
interactions between adjacent fermion fields. We are interested in the
general structure in the interactions, which may be deeply investigated
with the knowledge of spectral graph theory.

We hope that applications of our toy models to phenomenological models 
will be explored in future.





\bibliographystyle{apsrev4-1}


\end{document}